\documentstyle[aps,twocolumn,prl,psfig]{revtex} %% for single page

\begin{document}
%\draft
%\preprint{HEP/123-qed}
\title{Evidence for Static Magnetism in the Vortex Cores
of Ortho-II YBa$_2$Cu$_3$O$_{6.50}$ }

\author{
R.I.~Miller$^1$, R.F.~Kiefl$^{1,2,3}$, J.H.~Brewer$^{1,2,3}$,
J.E.~Sonier$^{2,4}$, J.~Chakhalian$^1$, S.~Dunsiger$^{1,*}$,
G.D.~Morris$^{2,*}$, A.N.~Price$^1$, D.A.~Bonn$^{1,3}$,
W.H.~Hardy$^{1,3}$, R.~Liang$^{1,3}$ }
%A. A.  Author and B. B. Author\cite{byline}\\
%Lines break automatically or can be forced
%with $\backslash\backslash$}
\address{$^1$
Dept. of Physics \& Astronomy, University of British Columbia,
Vancouver, Canada}
\address{$^2$ TRIUMF, Vancouver, Canada}
\address{$^3$ Canadian Institute for Advanced Research }
\address{$^4$
Simon Fraser University, Department of Physics \& Astronomy,
Burnaby, Canada}
\address{$^*$ present address:
Los Alamos National Laboratory, Los Alamos, New Mexico, USA}

\date{\today}
\maketitle
\begin{abstract}
Evidence for static alternating  magnetic fields in the vortex
cores of underdoped YBa$_2$Cu$_3$O$_{6+x}$ is reported. Muon spin
rotation measurements of the internal magnetic field distribution
of the vortex state of YBa$_2$Cu$_3$O$_{6.50}$ in applied fields
of $H \! = \! 1$ T and $H \! = \! 4$ T reveal a feature in the
high-field tail of the field distribution which is not present in
optimally doped YBa$_2$Cu$_3$O$_{6.95}$ and which fits well to a
model with static magnetic fields in the vortex cores. The
magnitude of the fields is estimated to be 18(2) G and decreases
above $T \! = \! 10$ K. We discuss possible origins of the
additional vortex core magnetism within the context of existing
theories.
\end{abstract}

\narrowtext

%%%%%  start of paper...
The application of a large magnetic field to a superconductor
drives part of the sample, the vortex cores, into a ``normal"
state.  While the physics of these vortex cores in conventional
superconductors is generally thought of as metallic, the vortex
cores of high temperature superconductors may offer insight into
more unusual low temperature properties. A new class of theories
has predicted that magnetism may be induced near/inside the cores
of vortices by the application of a magnetic
field\cite{zhang97}-\cite{demler01}. For example in Zhang's
$SO(5)$ theory, static antiferromagnetism (AF) should appear in
regions where the superconducting (SC) order parameter is
suppressed, and as a consequence, vortices in underdoped
superconductors should be magnetic\cite{zhang97,arovas97}.
Similarly, Lee and Wen have predicted a staggered flux phase (SFP)
of orbital currents in the vortex cores of underdoped
cuprates\cite{wen96,lee00}.  Below $T_c$, the system is locally a
$d$-wave superconductor away from the vortex cores, but has a SFP
``frozen'' inside the vortex cores. Consequently, they predict the
appearance of quasi-static alternating magnetic fields of order
$10$ G in the vortex cores. Finally, Zhu and Ting have developed
the Hubbard model with an on-site repulsion term and shown that
AF-like spin density waves (SDW) can appear in the vortex
cores\cite{zhu01},

The possible coexistence of magnetism and superconductivity in the
cuprates has been a key issue since the discovery of high-$T_c$
superconductors.  In zero field, microscopic coexistence of
superconductivity, magnetism, and spin-glass behavior has been
detected with $\mu$SR in the underdoped region near the boundary
between SC and AF\cite{brewer90}, in Ca-doped
YBa$_2$Cu$_3$O$_{6+x}$\cite{niedermayer98},
La$_{2-\delta}$Sr$_{\delta}$CuO$_4$\cite{niedermayer98,wakimoto01},
and underdoped and optimally doped
YBa$_2$Cu$_3$O$_{6+x}$\cite{jeff01,jeff01a}. There is also recent
experimental evidence from neutron scattering for anomalous
magnetism  in underdoped
YBa$_2$Cu$_3$O$_{6+x}$\cite{mook01,sidis01} in zero applied field.
Field-induced low-frequency magnetic fluctuations in
La$_{2-\delta}$Sr$_{\delta}$CuO$_4$ have been reported by
Lake\cite{lake01} and in YBa$_2$Cu$_3$O$_{6+x}$ by
Mitrovic\cite{mitrovic01}.  Vaknin \cite{vaknin00} found a
possible signature of AF cores in optimally-doped
YBa$_2$Cu$_3$O$_{6+x}$, while Katano report sharp incommensurate
neutron scattering peaks in La$_{2-\delta}$Sr$_{\delta}$CuO$_4$
that are enhanced in a magnetic field\cite{katano00}.

In this paper, we report a search for static magnetism in the
region of the vortex cores of underdoped Ortho-II
YBa$_2$Cu$_3$O$_{6.50}$. Significant improvements of the fits to
the muon spin precession signal in the vortex state are obtained
using a model of the vortex lattice with an additional alternating
magnetic field of $18(2)$ G, whose magnitude decreases away from a
vortex core center on the length scale of the coherence length. As
a control, similar fits were made on data from the conventional
superconductor NbSe$_2$ in which case no improvement in the fits
was observed, as expected.

Muons are an excellent probe of magnetism in superconductors. As
described elsewhere \cite{sonier_review}, the implanted spin
polarized muons stop randomly on the length scale of the vortex
lattice and precess at a rate proportional to the local magnetic
field, thus providing a direct measure of the local field
distribution $n(B)$. The observed asymmetric magnetic field
distribution in a superconductor is characteristic of a lattice of
magnetic vortices in a type-II superconductor.  In Ginzburg-Landau
(GL) theory, a vortex core's size is determined by the applied
magnetic field $H$ and the in-plane GL coherence length
$\xi_{ab}$, while the magnetic field decays away from the vortex
core over a length scale given by the GL $ab$ in-plane penetration
depth $\lambda_{ab}$. The magnetic field distribution $n(B)$ can
be calculated from the spatial distribution of the magnetic field
\cite{yaouanc}:

\begin{equation}
B(r)= {\Phi_0 \over S}(1-b^4) \sum_G e^{-i \vec{G}\cdot \vec{r}}
{uK_1(u) \over {1 + \lambda_{ab}^2  G^2}} \! , \label{eq:yaouanc}
\end{equation}
where $u^2=2 \xi_{ab}^2G^2(1+b^4)[1-2b(1-b)^2]$, $K_1(u)$ is a
modified Bessel function, $\vec{G}$ is a reciprocal lattice vector
of the vortex lattice,  $b=H/H_{c_2}$ is the reduced field,
$\Phi_0$ is the flux quantum, and $S$ is the area of the reduced
unit cell for a hexagonal lattice.

A phenomenological model of $n(B)$ that includes alternating
fields in the vortex cores can be created by adding a term to Eq.
\ref{eq:yaouanc}:

\begin{equation}
B'(r)= B(r) +
 (-1)^{(x+y)/a} M e^{-({r /
{\xi_{ab}}})^2} \! , \label{eq:moments}
\end{equation}
where $M$ is the amplitude of the alternating fields in the vortex
core that alternates in sign between neighboring crystal lattice
sites ($a$ is the distance between sites). Our model assumes that
the amplitude of the  fields in the vortex cores decays away from
the vortex core center over the same length scale $\xi_{ab}$ as
the SC order parameter increases. We note: 1) that Eq.
\ref{eq:moments} strictly holds only in the unit cell of the
vortex lattice, in the limit $\xi_{ab}/L_0 \! \ll 1$, where $L_0$
is the distance between vortex cores.  This ensures there is
negligible overlap between alternating fields of neighboring
vortices. 2) Because the vortex and crystal lattices are
incommensurate, the parameter $a$ has been varied between the
Cu-Cu distance in the planes and a much smaller value that allows
the muon to experience all alternating fields from $-M$ to $M$.
The results presented here are insensitive to such a change.

\begin{center}
\begin{figure}
\psfig{file=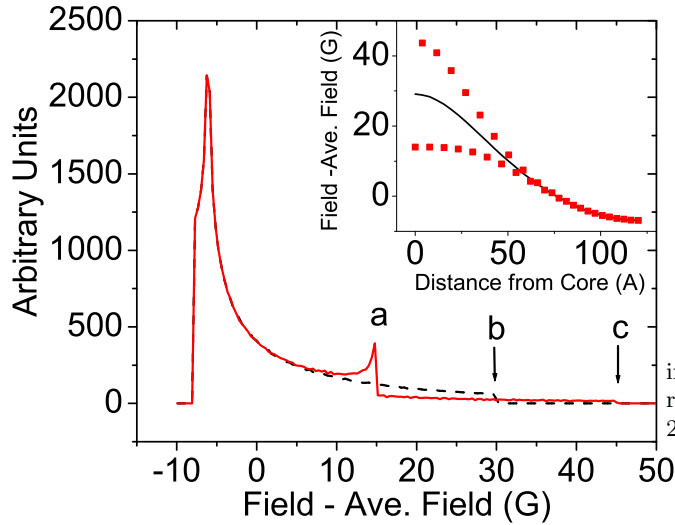,width=10cm} \caption[]{ Magnetic
field distribution $n(B)$ in superconductors in an applied field
of $H \! = \! 4$ T, with and without vortex core alternating
fields. Solid line: $n(B)$ calculated from Eq. \ref{eq:moments}
with alternating fields in the vortex cores ($M \! = \! 15$ G);
dashed line: field distribution without alternating fields
calculated from Eq. \ref{eq:yaouanc} with identical values for
$\xi$ and $\lambda$. Notice that significant changes occur only in
the high field tail. See text for further details. Inset: $B(r)$
along a line between two vortex cores. The solid line shows $B(r)$
without alternating fields, whereas the squares show $B(r)$ with
$M \! = \! 15$ G alternating fields in the vortex core. The
magnitude of the alternating fields is assumed to decay away from
the vortex core center on the length scale of $\xi_{ab}$. }
\label{fig1}
\end{figure}
\end{center}

Figure \ref{fig1} shows the effect of the vortex core alternating
fields on $n(B)$.  The dashed line shows $n(B)$ without
alternating fields ($M \! = \! 0$ G), for typical values  $\lambda
\! = \! 2450$ \AA~ and $\xi \! = \! 35$ \AA~ in an applied field
of $H \! = \! 4$ T. The high-field cutoff corresponding to the
magnetic field at the vortex core center is visible at $b$. The
modified $n(B)$ due to the introduction of the fields of magnitude
$M \! = \! 15$ G (solid line) shows a double step in the high
field tail at $a$ and $c$. The lower step (at $a$) is caused by
additional fields in the vortex core antiparallel to the applied
magnetic field $H$, whereas the upper step (at $c$) originates
from sites where the additional fields are aligned with the
applied field. The spatial variation of the magnetic field near
the vortex cores is shown in the inset. The line corresponds to
$B(r)$ from Eq. \ref{eq:yaouanc} without alternating fields, while
the squares show $B'(r)$ from Eq. \ref{eq:moments} with the
fields. The van Hove singularity in $n(B)$ at $a$ is caused by the
nearly flat field distribution in the inset along the lower line
of square points. These points correspond to fields aligned
opposite to the applied field.

With a model for $n(B)$, a theoretical function of the muon spin
precession signal $P(t)\! = \! P_{x} + iP_{y}$ is constructed from
Eq. \ref{eq:yaouanc} or \ref{eq:moments} by calculating the local
magnetic field at points on a lattice in the reduced unit cell and
summing the contributions from all sampling points:

\begin{equation}
P_x(t) = P(0) G(t) \int_{0}^{\infty} n(\omega) cos(\omega
t+\delta) d\omega, \label{eq:pol}
\end{equation}
where $\omega \! = \! \gamma_{\mu} B$, $G(t)$ is a gaussian
relaxation function that accounts for additional broadening due to
nuclear dipole moments and lattice disorder\cite{note1}, and $2
\pi \gamma_{\mu} \! = \! 135.54$ MHz/T.

In our fits to the muon spin precession signal, $n(B)$ parameters
$\xi_{ab}$, $\lambda_{ab}$, $M$, and $H$ were allowed to vary. It
has been shown previously that Eqs. \ref{eq:yaouanc} and
\ref{eq:pol} can be used to model the muon spin precession signal
in NbSe$_2$ \cite{sonier97a,miller} and YBa$_2$Cu$_3$O$_{6+x}$
\cite{sonier94,sonier97b} superconductors.  In this work,
constraints on the fits of Eq. \ref{eq:pol} were applied to the
range of allowed values of $\xi_{ab}$.  We find that at the
highest applied fields ($H \! = \! 4$ T), $\xi_{ab} \approx 25$
\AA, a value smaller than expected from previous $\mu$SR
measurements of YBa$_2$Cu$_3$O$_{6.60}$. Therefore, in the fits,
$\xi_{ab}$ is constrained to $\xi_{ab} \! > \! 35$ \AA~ at $H \! =
\! 4$ T. Unconstrained fits  yield {\it{larger}} values for the
alternating field magnitude.

The $\mu$SR experiments were performed on the M15 surface muon
channel at TRIUMF using a horizontal gas-cooled cryostat and a
high-field ($7.5$ T) high timing resolution spectrometer (BELLE).
All data were taken under conditions of field cooling, and high
statistics runs were made with approximately $30$ million muon
decay events for each value of $B,T$.  A mosaic of high purity
YBa$_2$Cu$_3$O$_{6.50}$ single crystals were used. They were grown
in BaZrO$_3$ crucibles, and detwinned below 200 C. Ortho-II
ordering was achieved by low temperature annealing for $7$ days
\cite{liang}.  The oxygen content of the sample was chosen due to
its stoichiometry and high degree of crystalline order, proximity
to the AF/SC boundary, and the oxygen ordering of the Ortho-II
phase.

$\mu$SR measurements performed in zero magnetic field in
YBa$_2$Cu$_3$O$_{6.50}$ show no internal magnetic fields.
Furthermore, the magnetic field dependence of the skewness and
second moment of the field distribution shows no signature of a
transition to a 2D pancake state\cite{pancake} at $ T \! = \! 2.2$
K, where $n(B)$ is no longer described by Eqs. \ref{eq:yaouanc}
and \ref{eq:moments}. In a field of $4$ T, an asymmetric field
distribution, the signature of a well-ordered lattice, is visible
up to $T \! = \! 30$ K. Further details of the vortex phase
diagram in YBa$_2$Cu$_3$O$_{6.50}$ will be published elsewhere.

\begin{center}
\begin{figure}
\psfig{file=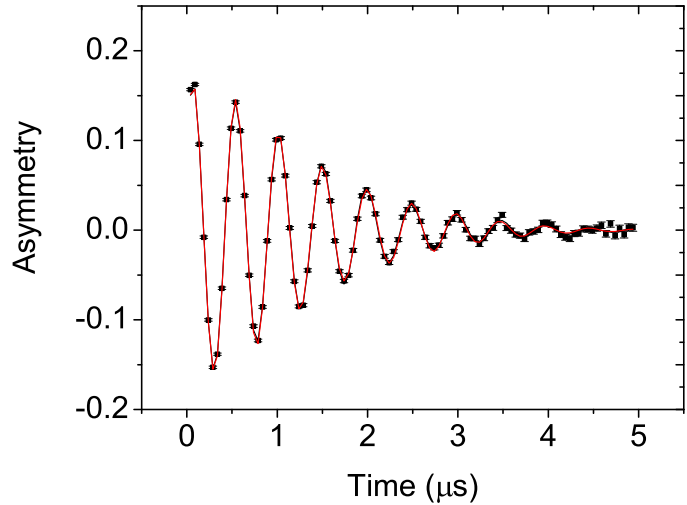,width=10cm} \caption[]{Muon
Polarization Signal P$_x$(t) in YBa$_2$Cu$_3$O$_{6.50}$ in an
applied field of $ H \! =\! 4$ T and at $T \! =\! 5$ K displayed
in a rotating reference frame. The line is a fit to P$_x$(t) using
Eqs. \ref{eq:moments} and \ref{eq:pol}.} \label{fig2}
\end{figure}
\end{center}

\begin{center}
\begin{figure}
\psfig{figure=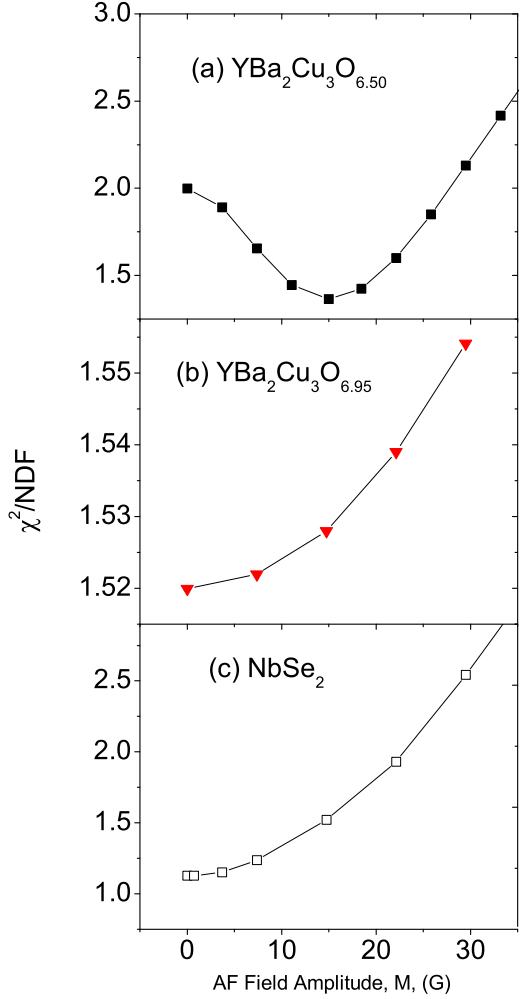,width=8cm} \caption[] { The best
$\chi^2$ per degree of freedom versus the amplitude of the
alternating field, M: (a) YBa$_2$Cu$_3$O$_{6.50}$ at $H \! = \! 4$
T and at $T \! =\! 5$ K (solid squares); (b)
YBa$_2$Cu$_3$O$_{6.95}$ at $ H \! = \! 6$ T (solid nablas); and
(c) for NbSe$_2$ at $H \! = \! 0.7$ T (open squares). The solid
curves are guides to the eye.} \label{fig3}
\end{figure}
\end{center}

Figure \ref{fig2} shows a fit to the muon spin precession signal
in YBa$_2$Cu$_3$O$_{6.50}$ assuming the phenomenological field
profile of Eq. \ref{eq:moments}. Notice the excellent fit of the
theoretical line to the data, and the small uncertainty in the
data due to the high statistics. The significant improvement in
the fits of our model with alternating magnetic fields is shown in
Figure \ref{fig3}. Fig. \ref{fig3}(a) shows the deep minimum in
$\chi^2/NDF$ (number of degrees of freedom) achieved by the
addition of fields of magnitude $M \! = \! 15$ G.  Figs.
\ref{fig3}(b) and (c) show that a similar effect is not found in
optimally doped YBa$_2$Cu$_3$O$_{6.95}$ (at $H \! = \! 6$ T) or in
the conventional superconductor NbSe$_2$ (at $H \! = \! 0.7$ T),
which is generally well described by BCS theory. The absence of
this effect in these materials indicates that the improvement of
the fits is not simply due to the addition of another parameter
and that the feature in YBa$_2$Cu$_3$O$_{6.50}$ is not present in
optimally doped YBa$_2$Cu$_3$O$_{6.95}$ or conventional
superconductors. The curves in Fig. \ref{fig3} are guides to the
eye.

\begin{center}
\begin{figure}
\psfig{file=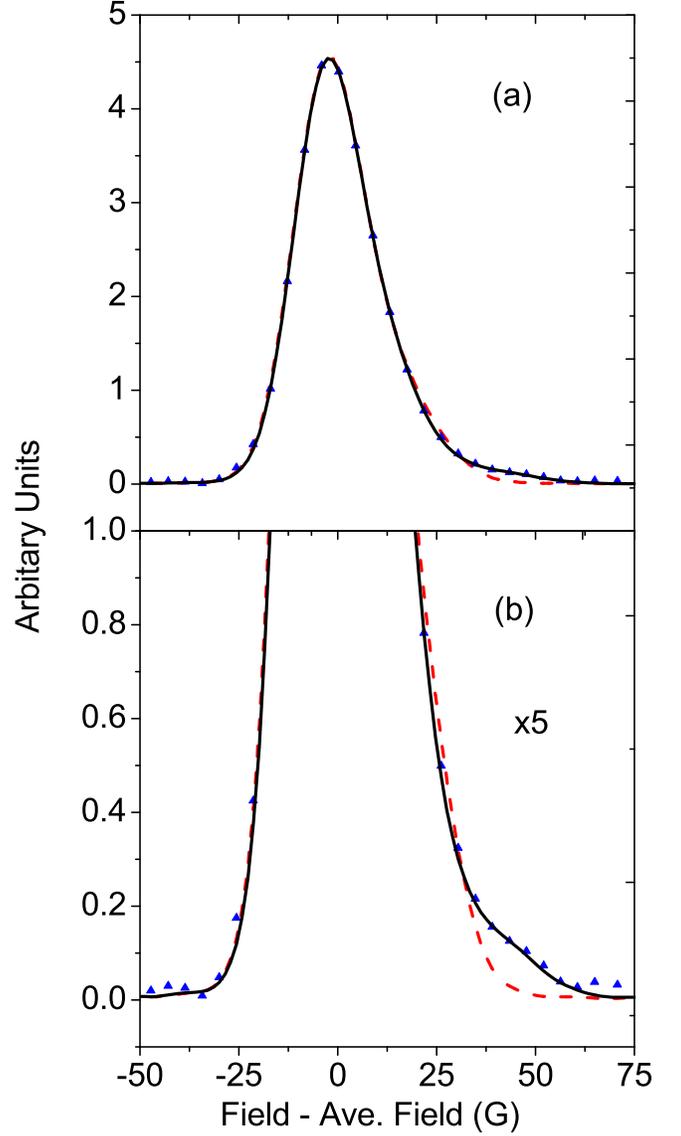,width=10cm} \caption[]{ (a) Fast
Fourier transform of $\mu$SR data (triangles) and fit (solid line)
in Fig. \ref{fig2} and of the best fit to Eqs. \ref{eq:yaouanc}
and \ref{eq:pol} without alternating fields (dashed line). (b) The
high-field region in greater detail.  Notice the improved fit in
the high field tail due to the alternating fields, corresponding
to the vortex core area.} \label{fig4}
\end{figure}
\end{center}

The improvements in the fit can be seen more clearly in the
frequency domain.  FFT's of data from Fig. \ref{fig2} and the
corresponding fits are shown in Fig. \ref{fig4}.  The triangles
are an FFT of the data, whereas the full (dashed) line is a FFT of
the fit with (without) vortex core  fields. Fig. \ref{fig4}(b)
shows the high field tail of the Fourier transforms in greater
detail. The improved fit is clearly visible in the high field
tail, despite the broadening introduced by Fourier transforming
finite data. Additional weight is present at fields higher than
the high field cutoff found using Eq. \ref{eq:yaouanc}.  Similar
improvements in the fits were found at other temperatures in
applied fields of $H \! = \! 4$ T and $H \! = \! 1$ T.  The
amplitude of the magnetic fields $M$ in the vortex core center
 decreases gradually with temperature above $T \! = \! 10$~K.
 Beyond $T\!=\!20$~K, no clear signature of the fields could be
 seen.

In our experiment, the muon is sensitive to the $c$-axis component
of the local magnetic field {\it at the muon site}.  Estimating
the size and direction of the effective magnetic moment
responsible for these alternating fields from these results is not
easy. However, Hsu {\it et al.} calculated that the geometric
factor due to the displacement between the muon site and the
center of a CuO plaquette would be of order $1$\cite{hsu90},
assuming the muon bonds to the oxygen above the CuO planes. Hence
measurement of $18$ G static fields at the muon site in the center
of the vortex core agrees quantitatively with theoretical
predictions\cite{lee00,arovas97}. Various other forms of the
alternating field term in Eq. \ref{eq:moments} have also been
tried. Replacing the gaussian function in the alternating field
term by an exponential yields a worse fit (although $\chi^2$ is
still a minimum at non-zero $M$), while fitting with a second
length scale for the magnetic field amplitude decay function
(instead of $\xi_{ab}$) provides only a marginal improvement in
the fit.

We note that there may be a number of other possible
interpretations of the anomalous $n(B)$ we have measured.  For
example, little is known about vortices in underdoped high-$T_c$
superconductors. The $d$-wave nature of the vortices or changes in
the pairing symmetry may affect $n(B)$. However, to our knowledge
no theories currently exist that predict a change in the
high-field tail due to such effects. Another possibility is that
the vortex lattice is characterized by two domains that coexist in
the sample, one a well-ordered vortex flux lattice and the other a
disordered flux lattice.  The high quality of the sample makes the
likelihood of such coexistence small\cite{liang}. Thus, we are led
to the conclusion that the most likely interpretation is the
presence of alternating fields in the vortex cores. As noted
above, numerous theories predict their existence, and experimental
evidence of fluctuating AF fields in the vortex cores of
La$_{2-\delta}$Sr$_{\delta}$Cu$_2$O$_4$ has been recently reported
in neutron scattering measurements. Our main result is precisely
the signature one would expect in the field distribution of a
superconductor with additional alternating magnetic fields in the
vortex cores.

In conclusion, we have shown that a phenomenological model of the
field distribution with alternating magnetic fields in the vortex
cores fits the muon spin precession signal in
YBa$_2$Cu$_3$O$_{6.50}$ in high magnetic fields significantly
better than a model without such vortex core  fields. We find the
 amplitude of these fields to be 18(2)$G$ at the muon site.
The origin of the anomalous magnetism reported here may be
consistent with Zhang's $SO(5)$ spin magnetism or various $t-J$
Hubbard models of orbital magnetism. Further measurements at
intermediate doping between $x \! =\! 0.5$ and $0.95$ are in
progress. This measurement is of significant importance to a class
of theories of high-T$_c$ superconductors that predict spin or
orbital magnetism in underdoped cuprates.

\acknowledgments

We wish to gratefully acknowledge discussions with Ian Affleck and
the help of Kim Chow, Mel Good, Bassam Hitti, Don Arseneau and Syd
Kreitzman and of the technical staff at TRIUMF.  Correspondence
should be addressed to R.I. Miller (miller@physics.ubc.ca).

%%%%%%%%%%%%%%%%%%%
%%%%%%%%%%%%%%%%%%%%%%%
\newpage

%%%%%%%%%%%%%%%%%%%
%%%%%%%%%%%%%%%%%%%%%%%
\newpage

%%%%%%%%%%%%%%%%%%%%
%%%%%%%%%%%%%%%%%%%%%%%

\newpage

%%%%%%%%%%%%%%%%%%%
%%%%%%%%%%%%%%%%%%%%%%%

%***
%***  E n d   o f   p a g e   6   o f   g a l l e y - m o d e   o u t p u t
%***

\end{document}